# Tuning hydrogen adsorption on graphene by gate voltage

Yuya Murata,[1] Arrigo Calzolari,[2] and Stefan Heun[1]

[1]NEST, Istituto Nanoscienze-CNR and Scuola Normale Superiore, Piazza San Silvestro 12, 56127 Pisa, Italy

[2]CNR-NANO Istituto Nanoscienze, S3 Center, Via Campi 213/a, 41125 Modena, Italy

**Abstract**

In order to realize applications of hydrogen-adsorbed graphene, a main issue is how to control hydrogen adsorption/desorption at room temperature. In this study, we demonstrate the possibility to tune hydrogen adsorption on graphene by applying a gate voltage. The influence of the gate voltage on graphene and its hydrogen adsorption properties was investigated by electrical transport measurements, scanning tunneling microscopy, and density functional theory calculations. We show that more hydrogen adsorbs on graphene with negative gate voltage (p-type doping), compared to that without gate voltage or positive gate voltage (n-type doping). Theoretical calculations explain the gate voltage dependence of hydrogen adsorption as modifications of the adsorption energy and diffusion barrier of hydrogen on graphene by charge doping.

# 1. Introduction

Chemisorption of hydrogen on graphene has been investigated due to its interest in modification of the electronic properties of graphene and application to hydrogen storage.[1] It has been theoretically and experimentally reported that a band gap opens upon hydrogen covalent bonding to graphene, transforming the hybridization of carbon atoms from $sp^2$ to $sp^3$,[2,3] and that the band gap increases with hydrogen coverage.[4] When graphene is used as a substrate for hydrogen storage, due to its high surface/mass ratio, the maximum gravimetric density which can be reached is 7.7 wt%,[2] a value larger than that of conventional high-pressure tanks.[5] In order to chemisorb hydrogen on graphene, hydrogen molecules must be dissociated to hydrogen atoms. Hydrogen molecules are dissociated by thermal energy which is given for example when they pass through a hot capillary. In case of graphene electronic devices with a band gap induced by hydrogen adsorption, exposure of atomic hydrogen on graphene will be done just once in the manufacturing process in a factory. On the other hand, in case of graphene hydrogen storage, exposure of atomic hydrogen must be done every time loading hydrogen, so it is not practical to thermally dissociate hydrogen molecules. In order to avoid this problem, methods to dissociate hydrogen molecules on graphene by electric field[6] or catalysts[7] have been proposed. The remaining issue in this context is to control hydrogen adsorption/desorption on graphene at room temperature. Theoretical calculations have predicted the possibility that hydrogen adsorption properties can be modified by electric field[8,9] and charge doping in graphene.[10,11,12] If this could be experimentally demonstrated, it would provide a simple mechanism to control hydrogen adsorption on graphene, useful for several applications. Electrical modification of the adsorption properties of molecules other than hydrogen on graphene have also been investigated.[13,14,15,16,17,18,19,20,21]

Here, using field effect transistor samples with graphene as its channel, we investigate the influence of a gate voltage applied to graphene on its hydrogen adsorption properties. The hydrogen adsorption was characterized by electrical transport measurements and scanning tunneling microscopy (STM) and spectroscopy (STS). Adsorption energy variation of hydrogen of graphene with doping was simulated by density functional theory (DFT) calculations. The experimental results show an increase of hydrogen adsorption on graphene with negative gate voltage. This was theoretically explained as an increase of adsorption energy of hydrogen on graphene due to p-type doping.

## 2. Methods

**2.A. Samples.** Graphene field effect transistor samples were prepared as follows. First, graphene flakes were transferred from a graphite piece to a $SiO_2$ (300 nm) / Si substrate by the scotch tape method. Two Au (45 nm) / Ti (5 nm) electrodes were fabricated on a selected graphene flake by electron beam lithography and thermal evaporation. Resist on samples for lithography was removed by soaking in acetone for 14 hours and rinsing in isopropanol for 3 min. Raman spectroscopy showed that graphene flakes are 2-3 layers thick, and do not have a D peak related to defects (not shown here). Samples were then introduced into an ultra-high vacuum chamber equipped with a hydrogen cracker (Tectra), a residual gas analyzer (Stanford Research Systems), and an STM (RHK technology). The pressure of the vacuum chamber was measured by an ionization gauge which was calibrated for $N_2$. The vacuum chamber has a base pressure of $3\times10^{-10}$ mbar, which rises to $\sim1\times10^{-9}$ mbar with the hydrogen cracker running. We have verified by residual gas analysis that the main contribution to this increase is due to hydrogen, and among the other gases, the one with the largest partial pressure was CO with $1.4\times10^{-10}$ mbar. Samples were heated at 500 K for 14 hours to further remove residues of resist from lithography. By STM it was confirmed that the graphene surface was clean.

**2.B. Hydrogenation.** For the hydrogenation experiments, the samples were exposed to atomic deuterium. While we generally refer to *hydrogen* throughout the paper, we specify the use of deuterium whenever relevant. $D_2$ gas was introduced from a gas bottle with purity higher than 99.8% to the vacuum chamber through a variable leak valve, and dissociated to atomic D by the hydrogen cracker. The hydrogen cracker consists of a tungsten capillary which is heated to 2000 K by electron bombardment. The sample was placed in front of the outlet of the hydrogen cracker, with a distance of 9 cm. The exposure time was controlled by a shutter between the hydrogen cracker and the sample. The atomic deuterium exposure was done by opening the variable leak valve until the ionization gauge read $\sim 1 \times 10^{-8}$ mbar. Considering the ion gauge sensitivity factor for $D_2$ of 0.29,[22] the partial pressure of $D_2$ was therefore $3.4 \times 10^{-8}$ mbar. Using a cracking efficiency of 100%, this corresponds to an atomic flux of $4.8 \times 10^{12}$ D atoms/(s cm$^2$), or 0.13% of the carbon density of graphene ($3.82 \times 10^{15}$/cm$^2$) per second.[23] After atomic deuterium exposure, the shutter and the variable leak valve were closed, the hydrogen cracker was turned off immediately, and the pressure of the vacuum chamber decreased to the base value within 1 min. Then the samples were characterized in the same UHV chamber by in-situ electric transport measurements and STM, without exposing the samples to air. Atomic deuterium exposure, electric transport measurements, and STM were performed at room temperature.

**2.C. Simulations.** First-principles total-energy-and-forces simulations are based on DFT, as implemented in the Quantum-Espresso[24] suite of codes. The exchange and correlation functional was expressed by using the van der Waals density functional (vdW-DF2) formulation,[25] and the spin degrees of freedom were treated within the local spin density approximation. The electron ion interactions were described by using ultrasoft pseudopotentials of Vanderbilt's type.[26] Single particle electronic wave functions (charge) were expanded in a plane wave basis set up to an energy cutoff of 28 Ry (280 Ry). All systems have been simulated by using periodically repeated supercells of size (17.04 × 17.22 × 15.00) Å$^3$, each including a graphene layer (112 C atoms), one or two H atoms, and a thick layer of vacuum (~15 Å) in the direction perpendicular to graphene, in order to avoid spurious interactions between adjacent replicas. Since the spatial extent of the graphene ripple observed experimentally (tens of nanometers)[27] is much larger than the typical size of systems that can be studied by DFT (few nanometers), here we assumed an ideal flat geometry as starting configuration for graphene before H adsorption. The Brillouin zone of the reciprocal lattice was sampled by a (6x6x1) grid of k-points, which explicitly includes the symmetry point K. Graphene doping is simulated by adding/removing (i.e. n/p-type) 0.05, 0.10, 0.15, and 0.20 electrons per cell, which spans the charge density range $\pm\,6.8 \times 10^{12}$ cm$^{-2}$. A jellium background is inserted to remove divergences in the charged cells. Atomic geometries of the separate subsystems (i.e., the molecule and the surface) were fully relaxed until forces were smaller than 300 meV/Å.

## 3. Experimental Results

The carrier concentration and therefore the charge doping in the graphene devices were controlled by a back gate. The induced charge is estimated by treating the graphene as one plate of a parallel plate capacitor, and the back gate (the highly doped Si substrate) as the other. Here, the dielectric between the two plates is the 300 nm-thick layer of SiO$_2$ with $\varepsilon \approx 3.9$, which gives a capacitance per area of $C$ = 11.5 nF/cm$^2$. The induced charge is then $n = C|V_{BG} - V_{CNP}|/e$, with $V_{BG}$ the applied back gate voltage, $V_{CNP}$ the back gate voltage at the charge neutrality point (CNP), for which the chemical potential of the device coincides with the Dirac point of graphene, and $e$ the elementary charge. The gate voltage for the electric transport measurements was swept with steps of 0.2 V and acquisition time of 0.5 s for each step. Such measurement takes about 5 min. Selected back gate sweeps are shown in Fig. 1. They show a maximum resistance value for minimum carrier concentration, i.e. at the CNP. Figure 1 shows that before atomic deuterium exposure, the CNP is located at about +10 V. This initial p-type doping of the graphene is probably due to charge transfer from the $SiO_2$ substrate and/or impurities between the graphene and the $SiO_2$ substrate.[28]

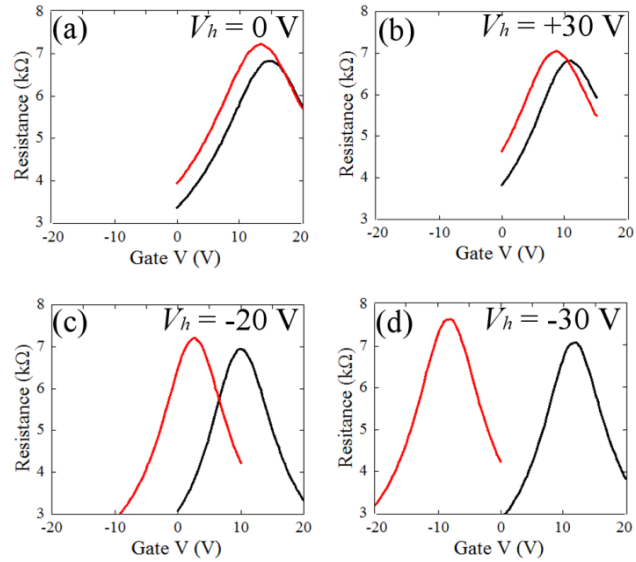

*FIG. 1. (Color online) Resistance of graphene as a function of gate voltage before (black) and after (red) atomic deuterium exposure. Atomic deuterium was exposed on graphene with (a) $V_h = 0$ V, (b) +30 V, (c) -20 V, and (d) -30 V for 5 s.*

Atomic deuterium was exposed to graphene for 5 s, while the graphene was grounded and a gate voltage $V_h$ was applied to the Si substrate. Before and after each hydrogenation step, we performed electric transport measurements in order to measure the change in charge density and resistance of graphene caused by deuterium adsorption. Figure 1 shows the resistance of graphene as a function of gate voltage before and after atomic deuterium exposure for different $V_h$ during atomic deuterium exposure. As already mentioned, in the initial condition the graphene is p-doped. After atomic deuterium exposure, for any value of $V_h$, $V_{CNP}$ had shifted to more negative values. This n-doping of graphene is attributed to deuterium adsorption, being hydrogen an electron donor for graphene.[29,30,31,32,33] The magnitude of the $V_{CNP}$ shift depends on $V_h$. The $V_{CNP}$ shift was found to be -20 V, -8 V, -1 V, and -1.5 V for $V_h$ = -30 V, -20 V, 0 V, and +30 V, respectively. This suggests that for negative $V_h$, more deuterium adsorbed on graphene, and more n-doping occurred, compared to the cases with zero or positive $V_h$. After atomic deuterium exposure, the resistance at $V_{CNP}$ increased slightly, indicating the presence of charge scattering centers induced by adsorbed deuterium.

We repeated these measurements on 4 different samples, in order to check the reproducibility of the $V_h$ dependence. Figure 2 shows the resulting $V_{CNP}$ shift as a function of $V_h$. All samples showed the same trend, independent of their thickness (2 or 3 layers), i.e. the $V_{CNP}$ shift is larger with negative $V_h$ compared to the cases with zero or positive $V_h$.

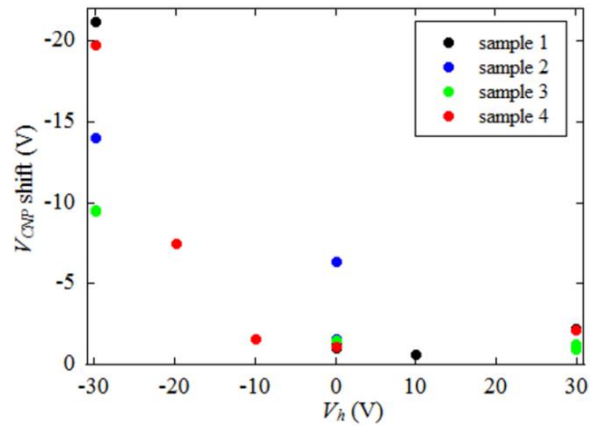

FIG. 2. *(Color online)* $V_{CNP}$ *shift by atomic deuterium exposure as a function of* $V_h$. *The colors correspond to data from 4 different samples.*

In order to evaluate the stability of the adsorbed deuterium, after atomic deuterium exposure with $V_h$ = +30 V and -30 V, the samples were kept in vacuum with gate voltage = 0 V, and the time evolution of $V_{CNP}$ was measured. The results are shown in Fig. 3. $V_{CNP}$ did not change much after atomic deuterium exposure with $V_h$ = +30 V, while it gradually shifted to slightly more positive values after atomic deuterium exposure with $V_h$ = -30 V. The $V_{CNP}$ shift to positive values may be due to desorption of adsorbed deuterium. In the case of $V_h$ = +30 V, the amount of adsorbed deuterium was small, therefore its desorption effect would be small, as well. However, even in the case of $V_h$ = -30 V, the $V_{CNP}$ shift in vacuum was only +2 V after 1 hour, which is much smaller than the $V_{CNP}$ shift due to the initial deuterium adsorption, which was -20 V. Thus, the results of Figs. 1 and 2 are not significantly affected by possible desorption effects. From Fig. 3 we can also deduce that the transport measurements have a negligible effect, if any, on the sample, because the variation in $V_{CNP}$ position between two consecutive back gate sweeps is less than 1 V, again much smaller than the measured $V_{CNP}$ shift of -20V due to the initial deuterium adsorption. In conclusion, under the experimental conditions employed here, deuterium desorption can be neglected, and the adsorbed deuterium can be considered stable at room temperature.

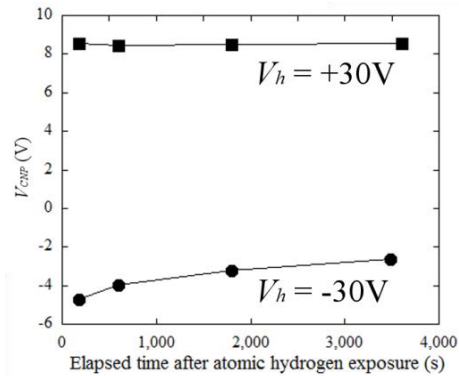

FIG. 3. Time evolution of the $V_{CNP}$ position after atomic deuterium exposure with (square) $V_h$ = +30 V and (circle) -30 V. The sample was kept in vacuum with gate voltage = 0 V.

The changes in the $V_{CNP}$ and resistance of graphene by atomic deuterium exposure can be reset by heating the sample at 600 K for 2 hours. This indicates thermal desorption of the adsorbed deuterium. Thermal desorption spectroscopy has shown that 600 K is high enough to desorb hydrogen dimers from graphite.[34,35] This suggests that the changes in $V_{CNP}$ and resistance of graphene by atomic deuterium exposure with gate voltage are reversible and are not due to any irreversible variation of the graphene structure such as creation of carbon vacancies.

We measured STM images of the graphene devices before and after atomic deuterium exposure. The results are shown in Fig. 4. The honeycomb lattice of graphene and the random corrugation of the $SiO_2$ substrate with an amplitude of approximately 1 nm and a periodicity of 10 nm were observed. However, even on the sample exposed to atomic deuterium with $V_h$ = -30V, shown in Fig. 4(b), we could not find any structure which we could clearly attribute to deuterium. This is not a problem of the tip condition, since the graphene lattice was well resolved. A possible reason for this is that the adsorbed deuterium cannot be distinguished from the large corrugation of the $SiO_2$ substrate. In fact, hydrogen atoms adsorbed on graphene on a SiC(0001) substrate, which has smaller and regular corrugation with an amplitude of 0.04 nm and a periodicity of 1.9 nm,[36] were observed as protrusions with a height of only 0.1 nm and a width of 0.2 nm.[3] Features due to instabilities of the tunnel junction like the horizontal bright lines in Fig. 4(b) might be caused by dragging deuterium on graphene by the STM tip.

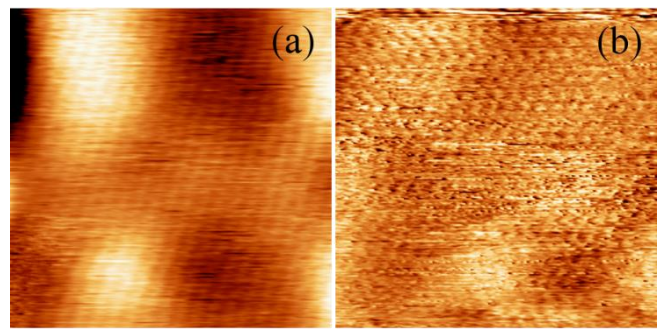

FIG. 4. (Color online) STM images on graphene (a) before and (b) after atomic deuterium exposure with $V_h$ = -30 V. (a) Bias voltage = 0.1 V, tunneling current = 0.16 nA, scan size = 5 nm × 5 nm, (b) bias voltage = 0.5 V, tunneling current = 0.03 nA, scan size = 5 nm × 5 nm.

Although we could not resolve individual adsorbed deuterium atoms by STM, we were able to detect changes in the electronic structure of the graphene upon deuterium adsorption by STS.

Figure 5(a) shows STS dI/dV spectra before and after atomic deuterium exposure with various $V_h$. The dI/dV signal was taken by a lock-in amplifier with a modulation voltage of 15 mV and a frequency of 921 Hz. The tip-sample distance was defined by a bias voltage of 0.4 V and a tunneling current of 0.5 nA. On each surface, at 10 different random positions, 20 spectra were taken and averaged. The spectra before and after atomic deuterium exposure with $V_h = 0$ V are almost the same. For the case of $V_h = +30$ V, the spectrum is slightly broader than the former two. On the other hand, for the case of $V_h = -30$ V, the spectrum indicates a relatively smaller density of states and a shift to negative energy. This suggests that with negative $V_h$, more deuterium adsorbed on graphene compared to the cases with zero or positive $V_h$, and adsorbed deuterium induced a band gap and a n-type doping. This is consistent with the electric transport measurements.

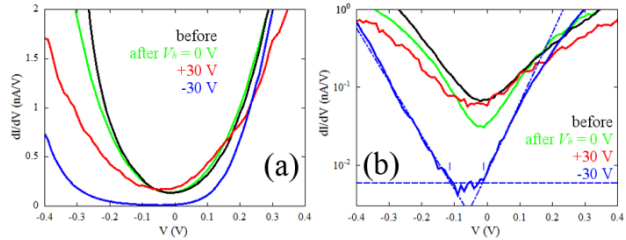

*FIG. 5. (Color online) (a) Averaged STS dI/dV spectra from graphene before (black) and after atomic deuterium exposure with (green) $V_h = 0$ V, (red) +30 V, and (blue) -30 V. Note the linear y-scale. (b) dI/dV spectra on a logarithmic scale. The blue dashed line is the average noise floor of the curve for $V_h = -30$ V. The blue dash-dotted lines are linear fits to the band edges. The band onsets are indicated by the blue vertical bars.*

We estimated the value of the band gap following a procedure reported earlier.[37] In short, the logarithm of each spectrum was taken and its average noise floor and standard deviation were determined. $E_{VB}$ and $E_{CB}$ were defined as the energies at which the conduction band and valence band edges, respectively, approach the standard deviation of the average noise floor. Linear fits were made to the spectrum for energies between $E_{VB}$ and $E_{VB} - 0.1$ eV, and for energies between $E_{CB}$ and $E_{CB} + 0.1$ eV. The band onsets were determined as the points where the linear fit lines intersect the average noise floor. The STS dI/dV spectra on a logarithmic scale are shown in Fig. 5(b). In the case of $V_h = -30$ V, the band gap width was estimated to be (0.14±0.05) eV, while the other spectra do not show a band gap. The relationship between band gap $E_{gap}$ and hydrogen coverage is approximately given by $E_{gap} = 3.8$eV (coverage/100%)$^{0.6}$.[4] For the hydrogenation experiments, the samples were exposed to deuterium, which induces a band gap in graphene similar to that by H adsorption. Consequently, it was suggested that the isotope effect on the band structure of graphene is weak.[38] Using the above relationship, the deuterium coverage was estimated to be (0.4±0.2)% from the band gap width measured by STS, i.e. 0.4% of the (graphene) surface C atoms bind to a D atom. The gravimetric density corresponding to a D coverage of 0.4% is 0.033 wt%. This is close to the atomic flux from the hydrogen cracker, 0.13%/s × 5 s = 0.64% (see section 2. Methods). This indicates that the sticking coefficient of deuterium under these experimental conditions is ~ 1. On the other hand, assuming a constant charge transfer per adsorbed deuterium, the sticking coefficient results to be 2-20 times smaller with $V_h = 0$ and +30 V. However, this value is still much larger than that of $10^{-4}$ reported for graphene on $SiO_2$ where residues of resist remained.[39] This confirms that the samples in this experiment were clean and free of resist.

On the other hand, the $V_{CNP}$ shift by the atomic deuterium exposure for this particular sample was -14 V. This corresponds to a variation in electron density by $1.0 \times 10^{12}$ cm$^{-2}$, considering the gate dielectric of $SiO_2$ with a width of 300 nm. From the deuterium coverage and the electron density, the electron transfer per adsorbed deuterium atom can be estimated to be 0.066 ±0.02 $e$. This is close to the value calculated by DFT, (0.06 $e$),[33] and supports our conclusion that the CNP shift and the band gap opening were induced by deuterium adsorption.

## 4. Simulations

In order to gain insight in the doping dependence of hydrogen adsorption on graphene, we carried out first principles simulations on both p-type and n-type doped graphene in the range $\pm 6.8 \times 10^{12}$ cm$^{-2}$, with respect to the neutral case. For each doping level we first optimized the atomic positions of the graphene layer (without H). In the absence of H, doping hardly modifies the original flat atomic structure of graphene. Furthermore, it does not break the Dirac cone degeneracy at the K point, but rather it imparts a rigid shift of the Fermi level in the valence (p-type) or conduction band (n-type). Then, starting from the relaxed positions of doped graphene, we included one H atom in the cell and optimized the interface. We obtain that H approaches the graphene layer and binds to a carbon atom (labeled $C_1$) in atop configuration (Fig. 6(a)). H adsorption induces an out-of-plane displacement ($\Delta z$) of the $C_1$ atom, which changes from pristine sp$^2$ to sp$^3$ configuration, in agreement with previous theoretical results.[40] While for all doping levels the final C-H distance is d(C-H) = 1.123 Å, the out-of-plane distortion of $C_1$ increases almost linearly with the charge density (Fig. 6(c)).

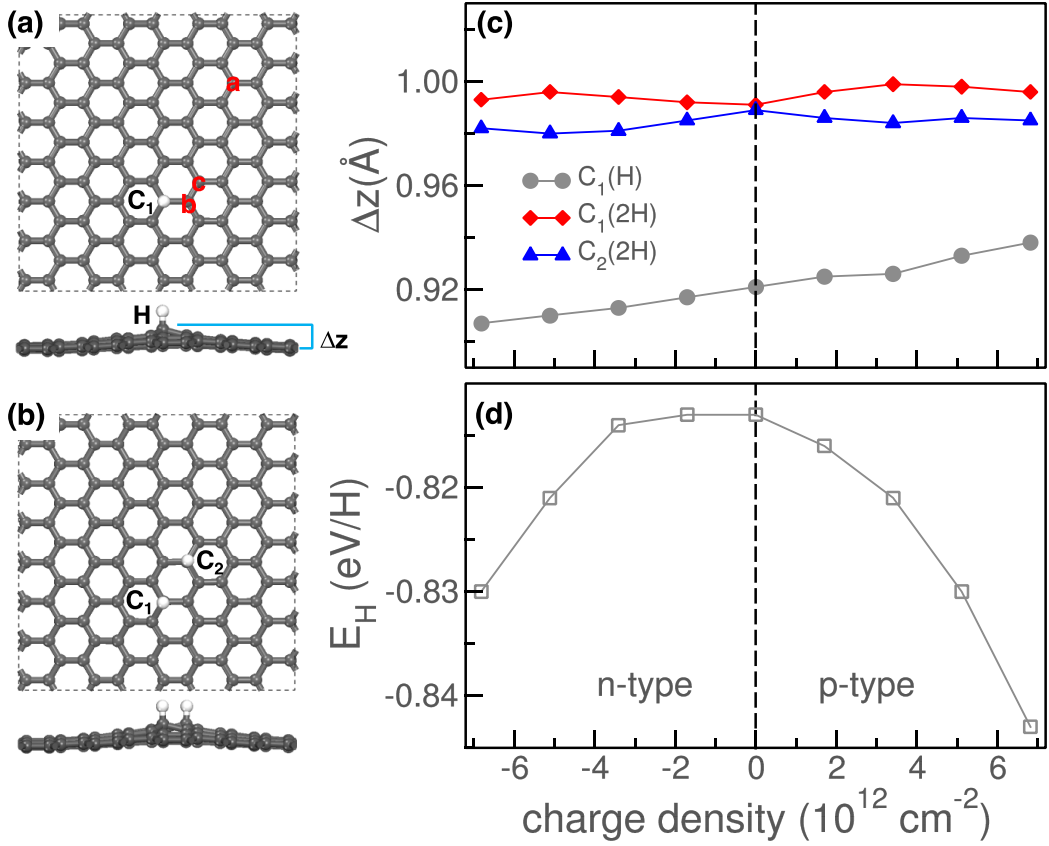

FIG. 6. Top and side view of (a) one and (b) two H atoms on graphene. (c) Out-of-plane displacement of Carbon atoms involved in C-H bonding, and (d) adsorption energy per hydrogen atom as a function of charge density in graphene. Dashed vertical line marks CNP, negative (positive) values correspond to n-type (p-type) doping, respectively.

The adsorption energy ($E_H$) of H on graphene varies as a function of graphene doping, as shown in Fig. 6(d). In particular, $E_H$ increases in absolute value as the amount of free charge (both electrons and holes) is increased, while for a fixed absolute value of free charge, the absolute value of $E_H$ is systematically higher for p-type doping than for n-type doping. $E_H$ has a maximum for the neutral configuration (lowest stability). In this case, the Fermi level is at the Dirac point (i.e. minimum density of states), and very few states are available for interacting with the incoming H states. Increasing doping in absolute value, the number of states at the Fermi level increases, which results in a higher coupling with the H and stabilizes the system. On the other hand, the asymmetry in the $E_H$ curve between p- and n-type doping can be ascribed to a not-trivial charge redistribution around the adsorption site, which differs with the charge density of the host. $E_H$ includes both a contribution associated to the formation of the C-H bond and a contribution due to the charge redistribution associated to the graphene distortion. Since the C-H distance results to be constant, the former contribution can be assumed to be the same for all configurations. Thus, the differences in $E_H$ are mostly attributed to the indirect effect that doping has on structural distortions, which break the ideal $D_{6h}$ symmetry of graphene, making the two carbon sublattices inequivalent and opening a bandgap.[2] We can conclude that H adsorbed as a monomer causes a structural distortion of graphene, whose amount and stability is modulated by doping, through a modification of the bonding/antibonding character of graphene states interacting with hydrogen.[10]

These results are in qualitative agreement with previous theoretical calculations which reported that with p-type doping of graphene the adsorption energy of hydrogen increases, and consequently the activation energy for desorption of hydrogen increases.[10,11,12] However, in previous calculations[10] the C-H distances were reported to change with doping, in contrast to our simulations. Nevertheless, two aspects have to be taken into account in this comparison: first, in those calculations the smallest charge density used was $2.5 \times 10^{13}$ cm$^{-2}$, i.e. one order of magnitude larger than that induced by $V_h = -30$ V in our experiment, $2.2 \times 10^{12}$ cm$^{-2}$, while in the present case, the charge density values considered in the DFT calculations are close to the experimental conditions for both types of doping. Secondly, in the previous work[10] the simulation cell was less than half of the present case (50 vs 112 C atoms per cell). Due to the applied periodic boundary conditions, this may change the possible lateral distortion of the graphene layer and, along with the different doping amount, may affect the final C-H distances.

Increasing coverage, adsorbed hydrogen atoms were typically observed as dimers or larger clusters in STM at room temperature.[3] This is because hydrogen monomers mostly desorb within minutes at room temperature, while hydrogen dimers are stable due to their higher adsorption energy.[41,42] We considered therefore the adsorption of a second H atom in the simulation cell. For each doping level, we started from the relaxed configurations described above (i.e. the distorted ones with the H monomers), and we added a second H atom, in three possible initial positions with respect to $C_1$, labelled **a**, **b**, and **c** in Fig. 6(a). When H atoms are far away (**a**), they do not interact and simply replicate the monomer configuration described above. On the contrary, when H atoms are very close together (**b**), they form a $H_2$ molecule that desorbs from graphene, a process which restores the initial flat configuration. This process is energetically not favored, and the adsorption energy is almost zero (-4 meV). Finally, if a second H is adsorbed relatively close to $C_1$ (**c**), it binds to the surface forming two interacting C-H dimers. Albeit the initial H is in a meta-position, after geometry optimization, H has diffused to a para-position for all considered doping levels (Fig. 6(b)). As for the monomer, also the carbon atom below the second H (labeled $C_2$ in Fig. 6) displaces out of plane in a $sp^3$ configuration, further distorting the carbon ring. For all charge density levels, the distances $d(C_1\text{-H}) = d(C_2\text{-H}) = 1.119$ Å, i.e. shorter than for the monomer case, while the distance $d(C_1\text{-}C_2) = 2.899$ Å is larger than in ideal graphene (d=2.840 Å). The out-of-plane displacement $\Delta z$ is systematically higher for both $C_1$ and $C_2$ than for the monomer case (Fig. 6(c)), with $C_1$ larger than $C_2$, i.e., the dimer is slightly asymmetric. Furthermore, the values of $C_1$ and $C_2$ are approximately independent of doping, see Fig. 6(c). The formation of dimers is energetically more stable than monomers by $\Delta E_H \sim 0.5$ eV per hydrogen atom. Since the bonding C-H lengths do not change with charge density, we can associate the extra gain $\Delta E_H$ to the more extended distortion of the graphene layer upon dimer formation. The structural contributions to $E_H$ are one order of magnitude larger than the doping ones, thus the effect of doping is no more as evident as for the monomer case: $E_H$ does not follow a clear trend with doping and has the same value $E_H = -1.359$ eV/H for all doping levels.

## 5. Discussion

We now discuss our results underlining the complementarity of the adopted approaches. Fig. 6 showed that p-type doping decreases $E_H$ of hydrogen monomers (increases $|E_H|$). This increases the residence time of hydrogen monomers on p-type graphene, which in turn increases the probability to form hydrogen dimers, for example with other hydrogen atoms directly impinging from the gaseous phase.

Besides, we consider also the case of diffusion of a monomer on the surface until it hits another monomer and forms a dimer. Previous theoretical calculations have reported that with p-type doping of graphene, the activation energy for diffusion decreases.[10] Furthermore, based on this report, another theoretical simulation has shown that with p-type doping of graphene, hydrogen monomers diffuse and form hydrogen dimers at room temperature, rather than to desorb.[43] On the other hand, n-type doping increases the activation energy for diffusion,[10] so that the activation energy for diffusion of hydrogen monomers becomes comparable to or larger than that for desorption, and therefore, on neutral or n-type graphene, hydrogen monomers will not diffuse to form hydrogen dimers, but rather desorb.[44]

Combining the above two effects derived from the theoretical calculations, we can conclude that p-type doping promotes conversion from hydrogen monomers to dimers. This explains the experimental results. Hydrogen monomers adsorbed on p-type graphene (p-doping by negative gate voltage) are converted to dimers, and they remain bound to graphene even after $V_h$ is turned to 0V. The CNP shift and the band gap of graphene induced by hydrogen dimers are detected by the transport measurement and in the STS data shown in Figs. 1, 2, and 5. A hydrogen sticking coefficient of nearly 1 is experimentally observed, consistent with this model. On the other hand, not all hydrogen monomers adsorbed on neutral or n-doped graphene are converted to dimers. This results in a loss of hydrogen via desorption and leads to a smaller hydrogen sticking coefficient, as experimentally observed.

For our experiments, the samples were exposed to deuterium. It is well known that H with its lower mass desorbs and diffuses on graphene more easily than D, due to the difference of their zero point energies.[34,44,45,46] According to a previous theoretical calculation,[10] the difference of the zero point energies between H and D is almost independent of the charge doping in graphene. Therefore, the desorption and diffusion barriers of H are different from those of D, but their variations by charge doping are the same for H and D. For this reason, the interpretation of the experimental results by the theoretical calculations about the charge doping dependence of desorption and diffusion is qualitatively applicable both for H and D.

The clear dependence of hydrogen adsorption energy on doping implies that external perturbations, such as gate voltage, which change the charge density of graphene, can be used to tune the stability of adsorbed hydrogen on graphene.

## 6. Summary

In summary, we demonstrated the possibility to tune hydrogen adsorption on graphene by applying a gate voltage. The influence of the gate voltage to graphene on its hydrogen adsorption properties was investigated by electric transport measurements and STM. After atomic hydrogen exposure to graphene with negative gate voltage, the $V_{CNP}$ shifted by a larger amount as compared to the cases with zero or positive gate voltage. A band gap opening was observed only after atomic hydrogen exposure to graphene with negative gate voltage. These results indicate that more hydrogen adsorbs on graphene with negative gate voltage, compared to that with zero or positive gate voltage. Our theoretical calculations revealed that p-type doping to graphene increases the adsorption energy of hydrogen. The experimental results are explained by an increase of residence time of hydrogen monomers on graphene and a consequent increase of conversion from monomers to stable dimers, by a p-type doping of graphene induced by a negative gate voltage.

**Acknowledgements**

Useful discussions with Valentina Tozzini and Richard Martel are gratefully acknowledged. We thank Fabio Beltram for his continuous support. The work received funding from the European Union's Horizon 2020 research and innovation program under grant agreement No. 696656 GrapheneCore1. Financial support from the CNR in the framework of the agreements on scientific collaborations between CNR and CNRS (France), NRF (Korea), and RFBR (Russia) is also acknowledged. We further acknowledge funding from the Italian Ministry of Foreign Affairs, Direzione Generale per la Promozione del Sistema Paese (agreement on scientific collaboration between Italy and Poland). S. H. thanks Scuola Normale Superiore for support, project SNS16_B_HEUN 004155.